\documentstyle[psfig]{mn}

\newif\ifAMStwofonts
\def\cm3{\thinspace\hbox{$\hbox{cm}^{3}$}}
\def\cm2{\thinspace\hbox{$\hbox{cm}^{2}$}}

\def\dyncm2{\thinspace\hbox{$\hbox{dyn}\thinspace\hbox{cm}^{-2}$}}

\def\C2{$\lambda\lambda$\thinspace\hbox{8542,8662}~\AA}

\def\lam{$\lambda$}

\def\etal {{\sl et\nobreak\ al.\ }}
\ifoldfss
  \ifCUPmtlplainloaded \else
    \NewTextAlphabet{textbfit} {cmbxti10} {}
    \NewTextAlphabet{textbfss} {cmssbx10} {}
    \NewMathAlphabet{mathbfit} {cmbxti10} {} % for math mode
    \NewMathAlphabet{mathbfss} {cmssbx10} {} %  "   "    "
  \fi
  \ifAMStwofonts
    \ifCUPmtlplainloaded \else
      \NewSymbolFont{upmath} {eurm10}
      \NewSymbolFont{AMSa} {msam10}
      \NewMathSymbol{\upi}     {0}{upmath}{19}
      \NewMathSymbol{\umu}     {0}{upmath}{16}
      \NewMathSymbol{\upartial}{0}{upmath}{40}
      \NewMathSymbol{$\leq$slant}{3}{AMSa}{36}
      \NewMathSymbol{\geqslant}{3}{AMSa}{3E}
           \let\oldleq=$\leq$
           
      \let$\leq$=$\leq$slant \let\le=$\leq$slant
      \let\geq=\geqslant 
    \fi
  \fi
\fi % End of OFSS

\ifnfssone
  \newmathalphabet{\mathit}
  \addtoversion{normal}{\mathit}{cmr}{m}{it}
  \addtoversion{bold}{\mathit}{cmr}{bx}{it}
  \newmathalphabet{\mathbfit} % math mode version of \textbfit{..}
  \addtoversion{normal}{\mathbfit}{cmr}{bx}{it}
  \addtoversion{bold}{\mathbfit}{cmr}{bx}{it}
  \newmathalphabet{\mathbfss} % math mode version of \textbfss{..}
  \addtoversion{normal}{\mathbfss}{cmss}{bx}{n}
  \addtoversion{bold}{\mathbfss}{cmss}{bx}{n}
  \ifAMStwofonts
    \ifCUPmtlplainloaded \else
      \UseAMStwoboldmath
      \makeatletter
      \new@mathgroup\upmath@group
      \define@mathgroup\mv@normal\upmath@group{eur}{m}{n}
      \define@mathgroup\mv@bold\upmath@group{eur}{b}{n}
      \edef\UPM{\hexnumber\upmath@group}
      \new@mathgroup\amsa@group
      \define@mathgroup\mv@normal\amsa@group{msa}{m}{n}
      \define@mathgroup\mv@bold\amsa@group{msa}{m}{n}
      \edef\AMSa{\hexnumber\amsa@group}
      \makeatother
      \mathchardef\upi="0\UPM19
      \mathchardef\umu="0\UPM16
      \mathchardef\upartial="0\UPM40
      \mathchardef$\leq$slant="3\AMSa36
      \mathchardef\geqslant="3\AMSa3E
           \let\oldleq=$\leq$
           
      \let$\leq$=$\leq$slant \let\le=$\leq$slant
      \let\geq=\geqslant 
    \fi
  \fi
\fi % End of NFSS release 1

\ifnfsstwo
  \DeclareMathAlphabet{\mathbfit}{OT1}{cmr}{bx}{it}
  \SetMathAlphabet\mathbfit{bold}{OT1}{cmr}{bx}{it}
  \DeclareMathAlphabet{\mathbfss}{OT1}{cmss}{bx}{n}
  \SetMathAlphabet\mathbfss{bold}{OT1}{cmss}{bx}{n}
  \ifAMStwofonts
    \ifCUPmtlplainloaded \else
      \DeclareSymbolFont{UPM}{U}{eur}{m}{n}
      \SetSymbolFont{UPM}{bold}{U}{eur}{b}{n}
      \DeclareSymbolFont{AMSa}{U}{msa}{m}{n}
      \DeclareMathSymbol{\upi}{0}{UPM}{"19}
      \DeclareMathSymbol{\umu}{0}{UPM}{"16}
      \DeclareMathSymbol{\upartial}{0}{UPM}{"40}
      \DeclareMathSymbol{$\leq$slant}{3}{AMSa}{"36}
      \DeclareMathSymbol{\geqslant}{3}{AMSa}{"3E}
           \let\oldleq=$\leq$
           
      \let$\leq$=$\leq$slant \let\le=$\leq$slant
      \let\geq=\geqslant 
    \fi
  \fi
\fi % End of NFSS release 2

\ifCUPmtlplainloaded \else
  \ifAMStwofonts \else % If no AMS fonts
    \def\upi{\pi}
    \def\umu{\mu}
    \def\upartial{\partial}
  \fi
\fi

\title[Empirical calibration of nebular abundances]
  {An empirical calibration of nebular abundances based on the sulphur emission 
lines.}
\author[Angeles I. D\'{\i}az \& Enrique P\'erez-Montero ]
  {Angeles I. D\'{\i}az and Enrique P\'erez-Montero\\
   Dpto. de F\'{\i}sica Te\'{o}rica, C-XI, Universidad Aut\'{o}noma de Madrid, Cantoblanco, 28049-Madrid, Spain}
\date{Accepted . Received 1998}
\pagerange{\pageref{firstpage}--\pageref{lastpage}}
\pubyear{1998}

\def\LaTeX{L\kern-.36em\raise.3ex\hbox{a}\kern-.15em
    T\kern-.1667em\lower.7ex\hbox{E}\kern-.125emX}

\begin{document}

\label{firstpage}

\maketitle

\begin{abstract}
 
We present an empirical calibration of nebular abundances based on the strong 
emission lines of [SII] and [SIII] in the red part of the spectrum through 
the definition of a sulphur abundance parameter $S_{23}$. This calibration 
presents two important advantages against the commonly used one based on the 
optical oxygen lines: it remains single-valued up to abundances close to solar 
and is rather independent of the degree of ionization of the nebula.

\end{abstract}


\section{Introduction}

The analysis of nebular spectra constitutes the best, and in some cases the 
only one, method for the determination of chemical abundances in spiral and
irregular galaxies, as well as in sites of recent star formation. The 
abundances of several elements like He, O, N and S can in principle be 
determined since strong emission lines of these elements, some of them  in their dominant 
ionization states, are present in the optical region of the spectrum. This 
requires knowledge of the electron temperature which can be obtained by 
measuring appropriate line ratios like \hbox{[OIII] $\frac{\lambda 4363}{\lambda 4959 + \lambda 5007}$}, \hbox{[NII] $\frac{\lambda 5755}{\lambda 6548 + \lambda 6584}$}, [OII]$\frac{ \lambda 7327}{\lambda 3727 + \lambda 3729}$, or [SIII]$\frac{\lambda 6312}{\lambda 9069 + \lambda 9532}$.

Unfortunately, these line ratios usually involve one intrisically weak line
which can be detected and measured with confidence only for the brighter and 
hotter objects and in many cases -- distant galaxies, low surface brightness 
objects, relatively low excitation regions -- they become too faint to be 
observed.

In these cases, an empirical method based on the intensities of the easily 
observable optical lines is widely used. The method, originally 
proposed by Pagel \etal\ (1979) and Alloin \etal\ (1979), relies on the 
variation of these lines with oxygen abundance. Pagel \etal (1979) defined an 
abundance parameter $R_{23} = \frac{[OII] \lambda 3727 +[OIII] \lambda 4959 + \lambda 5007}{H\beta}$ which increases with increasing abundance for abundances 
lower than about 20\% solar, and then reverses its behavior, decreasing with 
increasing abundance, since  above this value a higher oxygen abundance 
leads to a more effective cooling, the electron temperature gets lower and 
the optical emission lines get weaker. 

In principle, 
the calibration of the $R_{23}$ {\it versus} oxygen abundance relation can be 
done empirically in the low metallicity regime where electron temperatures can 
be derived directly, but requires the use of theoretical models for the so 
called high abundance branch. Several different calibrations have been made 
(Edmunds \& Pagel 1984; McCall \etal 1985; Evans \& Dopita 1986; Skillman 1989;
McGaugh 1991) as more observational data and more improved models have become 
available. However, two problems that are difficult to solve  still remain. The first 
one is related to the two-valued nature of the calibration, which can lead to 
important errors in the derived abundances. The second one concerns the 
dependence of $R_{23}$ on the degree of ionization of the nebula (see Skillman 
1989). $R_{23}$ also depends on density, but this can be considered as a second order effect for low density regions ($n_H$ $\simeq$ 100 cm$^{-3}$) which 
constitute the majority of the extragalactic population. These two facts, taken together, produce a large dispersion of the data 
for values of $logR_{23} \geq 0.8$ and $12+log(O/H) \geq 8.0$, with objects 
with the same value of $log R_{23}$ having actual abundances which differ by 
almost an order of magnitude. Unfortunately, a significant number of objects 
(about 40\%  of the observed HII galaxies; D\'\i az 1999) have $logR_{23} \geq 
0.8$ for which the calibration is most uncertain, and this percentage is even 
higher for HII regions in normal spiral galaxies.

%%%%%%%%%%%%%%%%%%%%%%%%%%%%%%%%%%%%%%%%%%%%%%%%%%%%%%%%%%%%%%%%%%%%%%%%%%%%%%%%
%
%     Figure 1  Oxygen abundance parameter vs oxygen abundance for the 
%               objects in Table 2 and the HII galaxies in Diaz 1998
%               Sulphur abundance parameter vs oxygen abundance for the 
%               objects in Table 2 
%
%
%%%%%%%%%%%%%%%%%%%%%%%%%%%%%%%%%%%%%%%%%%%%%%%%%%%%%%%%%%%%%%%%%%%%%%%%%%%%%%%%
\begin{figure}
%\begin{minipage}{180mm}
\psfig{figure=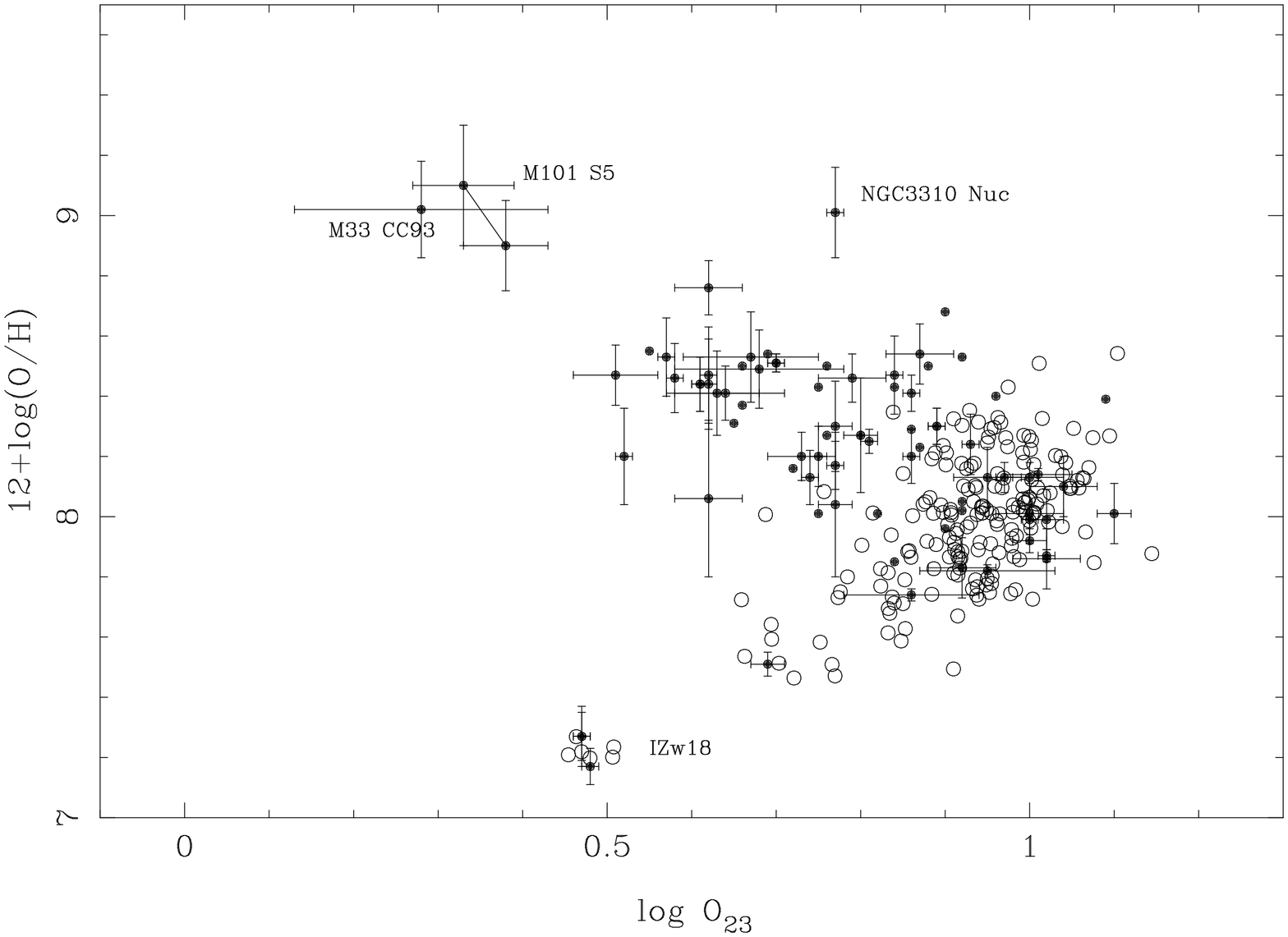,width=7cm,rheight=7.0cm,angle=270}
\vspace{12pt}
\psfig{figure=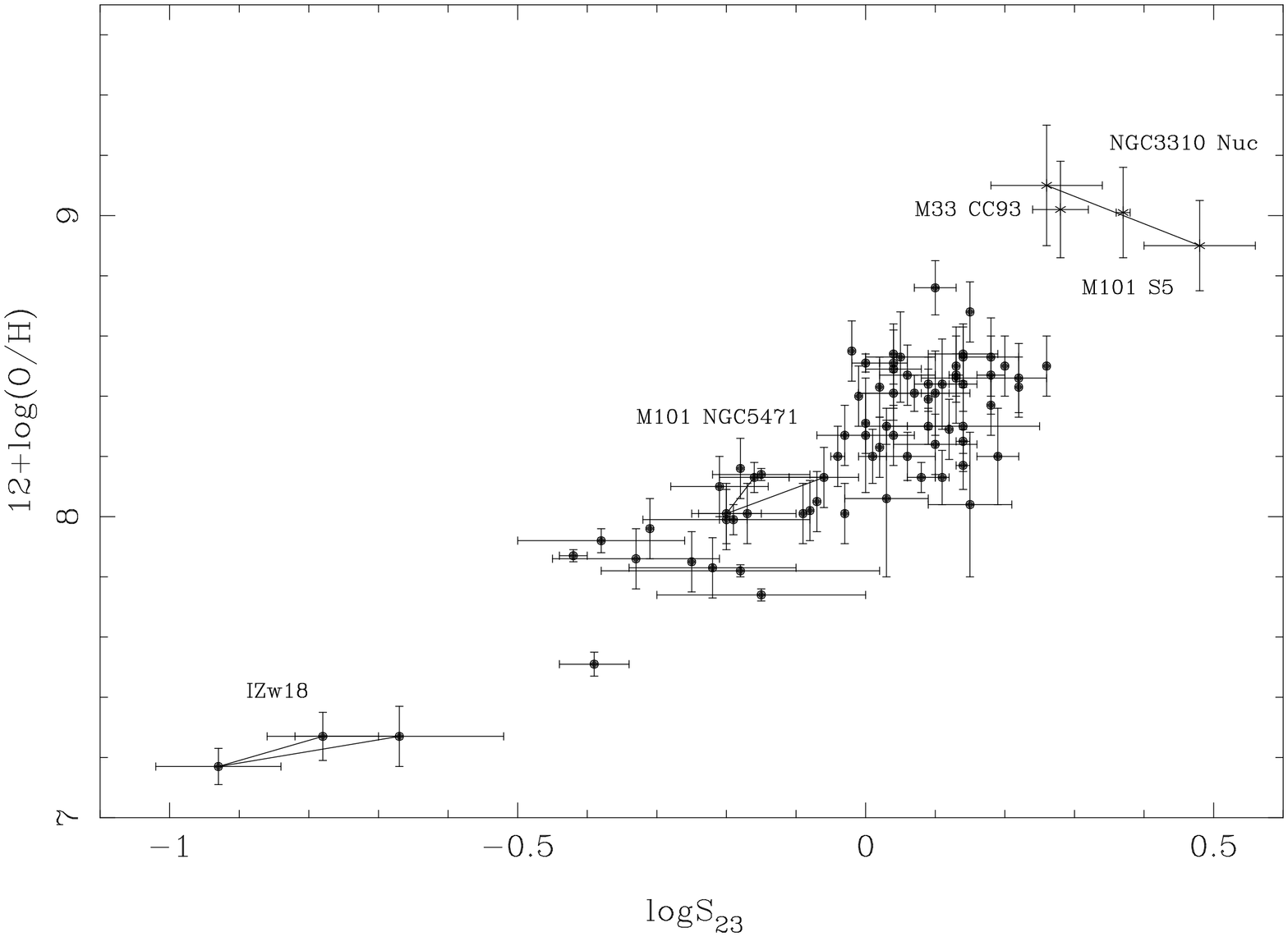,width=7cm,rheight=7.0cm,angle=270}
\caption{Oxygen abundance parameter (upper panel) and sulphur abundance parameter (lower panel) {\it vs} oxygen abundance for the objects in 
Table 2 and HII galaxies in D\'\i az (1998)}
%\end{minipage}
\end{figure}

Here we present an alternative abundance calibration based on the intensities 
of the sulphur lines: [SII] \lam\ 6716, \lam\ 6731 and [SIII] \lam\ 9069, \lam\
9532, through the use of the sulphur abundance parameter S$_{\rm 23}$ 
(V\'\i lchez \& Esteban 1996).
 
Spectroscopically, these lines are analogous to the optical oxygen lines 
defining $R_{23}$ but, due to their longer wavelengths, their contribution 
to the cooling of the nebula should become important at a somewhat lower temperature. Yet, the lower abundance of sulphur makes these lines less 
significant than the [OIII] lines as a contributor to cooling. On the other hand, the sulphur lines are less sensitive to temperature, therefore  the reversal in the relation between their intensities 
and the average nebular abundance is expected to occur at a higher metallicity 
and the relation should remain single valued longer.

From the observational point of view they present two important advantages: first, the lines are easily detected both in high and low excitation ionized 
regions  (D\'\i az {\it et al.} 1990) and second, they are less affected by reddening; moreover, they can be measured relative to nearby hydrogen recombination lines, H$\alpha$ in the case of the [SII] lines and Paschen lines
in the case of the [SIII] lines, which minimizes also any flux calibration 
uncertainties. These lines are accessible spectroscopically with CCD detectors up to a redshift of about 0.1. 
  
%%%%%%%%%%%%%%%%%%%%%%%%%%%%%%%%%%%%%%%%%%%%%%%%%%%%%%%%%%%%%%%%%%%%%%%%%%%%%%%%
%
%     Figure 2 Relation between logO23 (upper panel) and logS23 (lower 
%              panel with log([OII]/[OIII]).
%
%
%%%%%%%%%%%%%%%%%%%%%%%%%%%%%%%%%%%%%%%%%%%%%%%%%%%%%%%%%%%%%%%%%%%%%%%%%%%%%%%%
\begin{figure}
%\begin{minipage}{180mm}
\psfig{figure=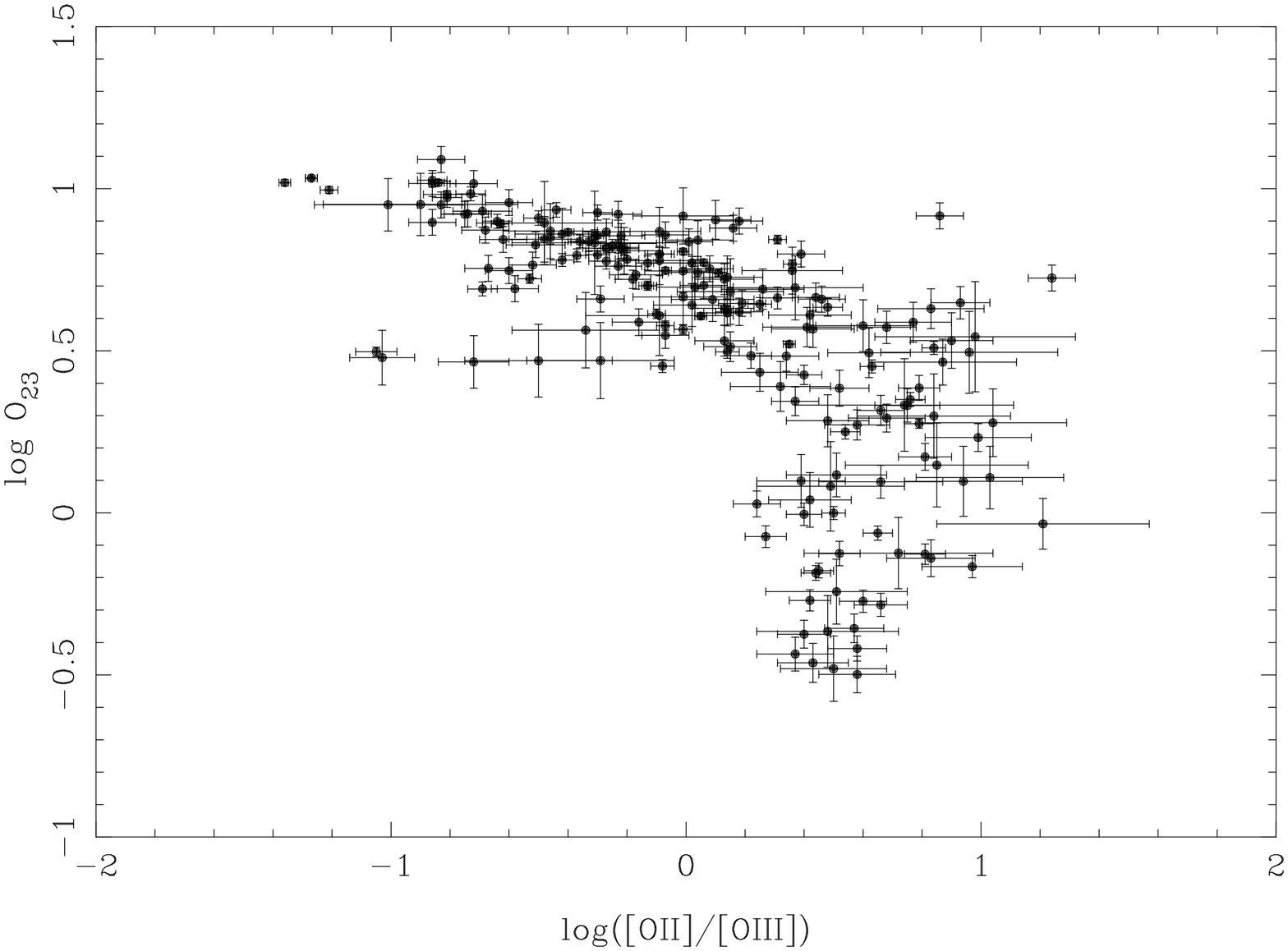,width=7cm,rheight=7.0cm,angle=270}
\psfig{figure=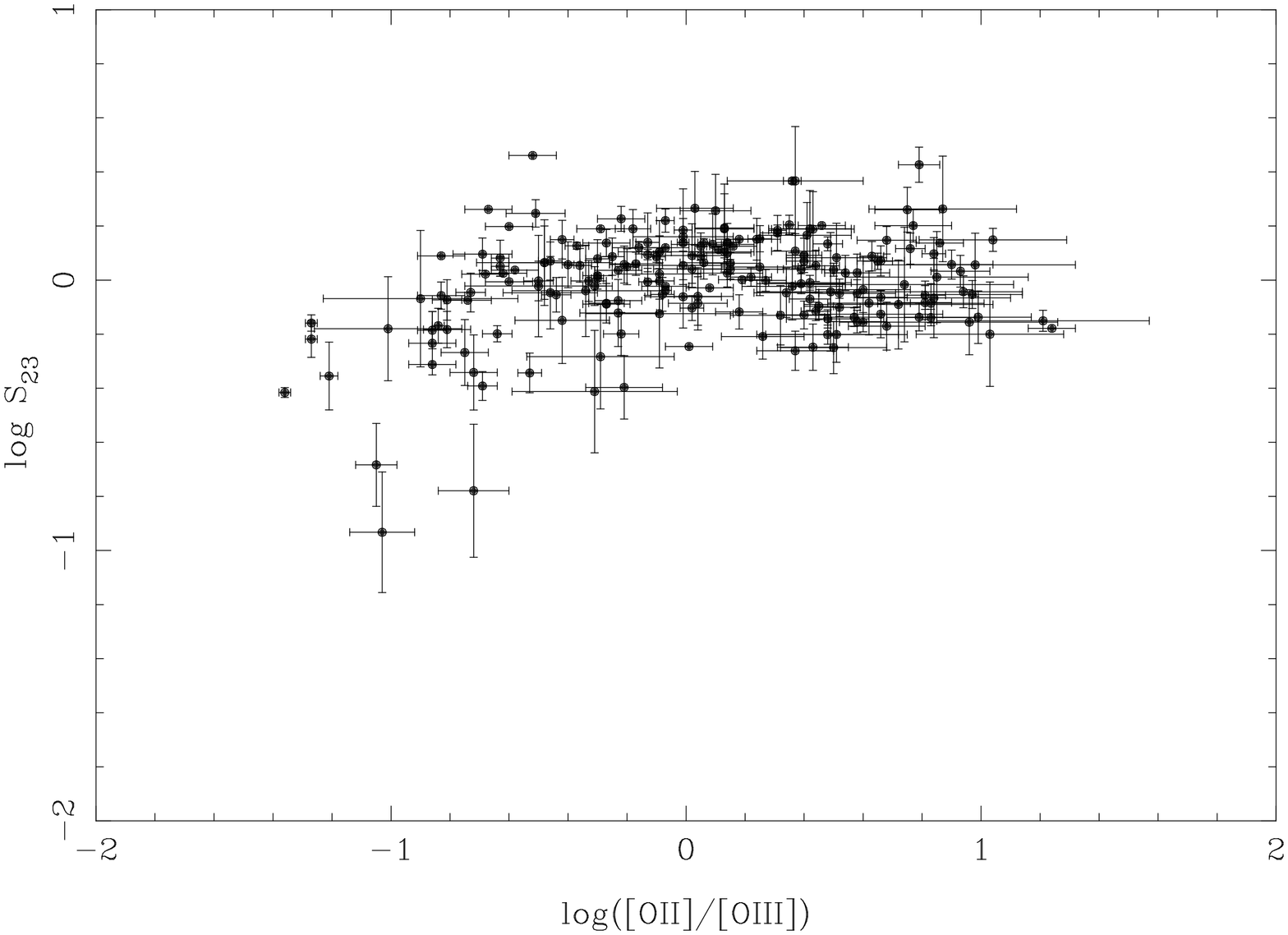,width=7cm,rheight=7.0cm,angle=270}
\caption{Relation between $log_{O23}$ (upper panel) and $logS_{23}$ (lower              panel with $log([OII]/[OIII])$, taken as ionization parameter indicator, for 
the objects in Table 1.}
%\end{minipage}
\end{figure}

\section{Observational Data}

In the last decade there has been a considerable increase in the amount of 
near infrared [SIII] emission line data. We have compiled from the literature 
all the HII region like objects for which measurements of these lines, together 
with optical emission lines exist. The sample includes HII regions in our own 
Galaxy, HII regions in spirals and irregular galaxies and HII galaxies. The 
data are presented in Table 1, where the logarithmic intensities of the 
emission lines with respect to H$\beta$ are given, together with their 
corresponding observational errors when quoted in the original reference. These 
references are numbered in column 3 and listed at the bottom of the table.
The observed intensities were corrected for reddening according to the derived
logarithmic extinction at $H\beta$ listed in the last column of the table and a 
standard reddening law. The sample comprises a total of 196 objects.

%%%%%%%%%%%%%%%%%%%%%%%%%%%%%%%%%%%%%%%%%%%%%%%%%%%%%%%%%%%%%%%%%%%%%%%%%%%%%%%%
%
%             Tabla 1 Intensidades de las lineas de emision
%
%%%%%%%%%%%%%%%%%%%%%%%%%%%%%%%%%%%%%%%%%%%%%%%%%%%%%%%%%%%%%%%%%%%%%%%%%%%%%%%%

\begin{table*}

 \begin{minipage}{180mm}

 \caption{Oxygen and Sulphur emission line intensities}

 \label{symbols}

 \begin{tabular}{|ll|c|cc|cc|cc|cc|c|}

  \bf Galaxy & \bf Region & \bf Ref. & \bf [OII] & \bf $\Delta$ &\bf [OIII] &\bf

$\Delta$ & \bf [SII] & \bf $\Delta$ & \bf [SIII] & \bf $\Delta$ & \bf c(H$\beta$)  \\ 

        &    &  & 3727+ 3729 &  & 4959 + 5007 & & 6716+6731 & & 9069+9532 & & \\                   

 \hline

\setcounter{table}{0}%

MWG &      NGC2467    &   1   &  0.49   &   ---   &   0.41  &    ---  &   -0.71 &     --- &    -0.13    &  ---  &    0.60  \\

 -- &    ETACAR     &   1   &  0.19   &   ---   &   0.48  &    ---  &   -0.54 &     --- &     0.10    &  ---  &    0.70  \\

 -- &     M17        &   1   &  0.05   &   ---   &   0.65  &    ---  &   -1.19 &     --- &     0.18    &  ---  &    1.32  \\

 -- &     M16        &   1   &  0.21   &   ---   &   0.28  &    ---  &   -0.65 &     --- &    -0.14    &  ---  &    1.12  \\

 -- &     M20        &   1   &  0.53   &   ---   &   0.07  &    ---  &   -0.38 &     --- &     0.07    &  ---  &    0.66  \\

 -- &     NGC3576    &   1   &  0.13   &   ---   &   0.65  &    ---  &   -0.87 &     --- &     0.44    &  ---  &    0.76  \\

 -- &     ORION 1    &   1   &  0.01   &   ---   &   0.59  &    ---  &   -1.38 &     --- &     0.02    &  ---  &    0.85  \\

 -- &     ORION 2    &   1   &  0.00   &   ---   &   0.67  &    ---  &   -1.32 &     --- &     0.25    &  ---  &    0.63  \\

 -- &     S209       &   19  &  0.08   &  0.03   &   0.61  &   0.01  &  -0.82  &   0.06  &  -0.52  &   0.08  &   3.08  \\

 -- &    S283       &   19   &  0.60   &  0.05   &  -0.33  &   0.05  &  -0.11  &   0.05  &  -0.52  &   0.08  &   1.44  \\

 -- &    S98        &   19   &  0.57   &  0.05   &  -0.26  &   0.13  &  -0.31  &   0.06  &  -0.48  &   0.09  &   2.35  \\

 -- &    S127       &   19   &  0.40   &  0.05   &   0.22  &   0.03  &  -0.76  &   0.04  &  -0.23  &   0.07  &   2.37  \\

 -- &    S128       &   19   &  0.32   &  0.05   &   0.59  &   0.01  &  -0.72  &   0.05  &  -0.20  &   0.08  &   2.25  \\

IC10      &  \# 2     &   4   &  0.16   &  0.05    &   0.85  &  0.05   &   -0.70 &    0.06 &    0.02     &   0.06  &   1.48  \\

SMC &   N80        &  1   &  0.54   &   ---   &   0.53  &    ---  &   -0.85 &     --- &    -0.37    &  ---  &   -1.00  \\

 --  &   N83        &  1   &  0.36   &   ---   &   0.63  &    ---  &   -0.52 &     --- &    -0.29    &  ---  &    0.12  \\

 --  &   N13        &  1   &  0.49   &   ---   &   0.72  &    ---  &   -1.01 &     --- &    -0.13    &  ---  &   -1.00  \\

 --  &   N32        &  1   &  0.70   &   ---   &  -0.54  &    ---  &   -0.46 &     --- &    -0.50    &  ---  &    0.15  \\

 --  &   N81        &  1   &  0.11   &   ---   &   0.85  &    ---  &   -0.93 &     --- &    -0.14    &  ---  &    0.04  \\

 --  &   N66        &  1   & -0.02   &   ---   &   0.84  &    ---  &   -0.97 &     --- &    -0.42    &  ---  &    0.17  \\

NGC295   & 1        &   21   &  0.44   &  0.04   &   0.45  &   0.09  &  -0.41  &   0.07  &  -0.32  &   0.15  &   0.50  \\

    --     & 2        &   21   &  0.56   &  0.03   &   0.52  &   0.09  &  -0.31  &   0.06  &  -0.42  &   0.16  &   0.57  \\

    --     & 3        &   21   &  0.52   &  0.04   &   0.61  &   0.10  &  -0.36  &   0.08  &  -0.21  &   0.18  &   0.60  \\

    --     & 6        &   21   &  0.24   &  0.04   &   0.72  &   0.08  &  -0.51  &   0.06  &  -0.07  &   0.16  &   0.43  \\

    --     & 7        &   21   &  0.59   &  0.05   &   0.23  &   0.12  &  -0.30  &   0.09  &  -0.35  &   0.16  &   0.54  \\

    --     & 8        &   21   &  0.40   &  0.04   &   0.31  &   0.10  &  -0.20  &   0.08  &  -0.14  &   0.14  &   0.39  \\

    --     & 9        &   21   &  0.29   &  0.09   &   0.77  &   0.14  &  -0.36  &   0.12  &  -0.14  &   0.18  &   0.40  \\

M33 &   CC93       &   20   &  0.24   &   ---   &  -0.80  &    ---  &   -0.06 &    0.05 &   -0.27     &   0.03  &   0.20  \\

 --  &   IC142      &   20   &  0.36   &  0.01   &   0.01  &   0.01  &   -0.32 &    0.07 &    0.05     &   0.02  &   0.24  \\

 --  &   NGC595     &   20   &  0.33   &  0.01   &   0.28  &   0.01  &   -0.60 &    0.04 &   -0.01     &   0.01  &   0.49  \\

 --  &   NGC595     &    4   &  0.33   &  0.01   &   0.28  &   0.01  &   -0.51

&    0.05 &    0.01    &   0.05  &   0.49  \\     

 --  &   MA 2       &   20   &  0.26   &  0.02   &   0.36  &   0.01  &   -0.71 &    0.02 &    0.01     &   0.02  &   0.10  \\

 --  &   NGC604     &   20   &  0.33   &  0.01   &   0.46  &   0.01  &   -0.43 &    0.04 &   -0.06     &   0.01  &   0.36  \\

 --  &   NGC604     &    4   &  0.33   &  0.02   &   0.46  &   0.01  &   -0.51

&    0.05 &   -0.17     &   0.05  &   0.36  \\

 --  &   NGC588     &   20   &  0.17   &  0.02   &   0.80  &   0.01  &   -0.72 &    0.11 &   -0.03     &   0.02  &   0.15  \\

 --  &   NGC588     &    4   &  0.17   &   0.02  &   0.80  &   0.01  &   -0.64 &    0.05 &   -0.01     &   0.07  &   0.15  \\         

 --  &   IC131      &    4   &  0.32   &  0.01   &   0.72  &   0.01  &   -0.57 &    0.10 &   -0.06     &   0.08  &   0.10  \\

NGC753 & 1        &  12   &  0.26   &  0.15    &   0.35  &  0.10  &  -0.20  &   0.15  &  -0.92  &   0.40   &   0.72  \\

    --     & 4        &  12   & -0.15   &  0.15    &   0.35  &  0.10   &  -0.13  &   0.15   &  -0.68  &   0.30   &   0.18  \\

    --    & 17        &  12   & -0.04   &  0.15    &  -0.53  &  0.10   &  -0.28  &   0.15   &  -0.42  &   0.30   &   0.85  \\

    --    & 20        &  12   &  0.00   &  0.15    &   0.29  &  0.10   &  -0.41  &   0.15   &  -0.88  &   0.30   &   0.35  \\

    --    & 23        &  12   &  0.06   &  0.15    &   0.40  &  0.10   &  -0.09  &   0.15   &  -1.00  &   0.30   &   0.99  \\

Mkn 600   &  -- &   4   &  0.22   &  0.04   &   0.94  &   0.04  &  -0.73  &   0.09  &  -0.57  &   0.17  &   0.66  \\

LMC &   N59A       &  1   &  0.26   &   ---   &   0.86  &    ---  &   -0.87 &     --- &    -0.07    &  ---  &    0.19  \\

 --  &   N44B       &  1   &  0.20   &   ---   &   1.03  &    ---  &   -0.74 &     --- &     0.02    &  ---  &    0.09  \\

 --  &   N55A       &  1   &  0.33   &   ---   &   0.56  &    ---  &   -0.96 &     --- &    -0.01    &  ---  &    0.10  \\

 --  &   N113D      &  1   &  0.68   &   ---   &   0.50  &    ---  &   -0.67 &     --- &     0.08    &  ---  &   -1.00  \\

 --  &   N127A      &  1   &  0.65   &   ---   &   0.49  &    ---  &   -0.73 &     --- &     0.06    &  ---  &   -1.00  \\

 --  &   N159A      &  1   &  0.11   &   ---   &   0.79  &    ---  &   -1.01 &     --- &    -0.02    &  ---  &    0.43  \\

 --  &   N214C      &  1   &  0.52   &   ---   &   0.59  &    ---  &   -0.37 &     --- &    -0.05    &  ---  &    0.00  \\

 --  &   N4A        &  1   &  0.13   &   ---   &   0.75  &    ---  &   -0.90 &     --- &    -0.03    &  ---  &    0.25  \\

 --  &   N79E       &  1   &  0.43   &   ---   &   0.24  &    ---  &   -0.53 &     --- &    -0.15    &  ---  &    0.16  \\

 --  &   N191A      &  1   &  0.86   &   ---   &   ---   &    ---  &   -0.43 &     --- &     0.00    &  ---  &   -1.00  \\

IIZw40    & --     &   4   & -0.26   &  0.02   &   1.006  &   0.004  &  -0.83  &   0.06  &  -0.34  &   0.07  &   1.23  \\

IIZw40     & --     &    2   & -0.26   &  0.02   &   1.006  &   0.004  &  -0.98 &   0.03  &  -0.23  &   0.03  &   1.23  \\

NGC2366 &NGC2363 &   4   & -0.24   &  0.02   &   0.97  &   0.01  &  -1.06  &   0.06  &  -0.450 &   0.14  &   0.25  \\

NGC2366   &NGC2363~A1   &  8  &  0.30   &  0.08   &   0.72  &   0.08  &  -0.52  &   0.08  &  -0.39  &   0.21  &   0.19  \\

      --  &NGC2363~A2   &  8   & -0.36   &  0.01   &   1.00  &   0.01  &  -1.17  &   0.01  &  -0.50  &   0.02  &   0.20  \\

       --   &NGC2363~A3   &  8   & -0.10   &  0.18   &   0.91  &   0.07  &  -0.83  &   0.08  &  -0.29  &   0.22  &   0.18  \\

       --  &NGC2363~A4   &  8   &  0.00   &  0.26   &   0.90  &   0.07  &  -0.60  &   0.11  &  -0.22  &   0.30  &   0.18  \\

\hline

  \end{tabular}

 \end{minipage}

 \end{table*}

\begin{table*}

 \begin{minipage}{180mm}

 \setcounter{table}{0}%

\caption{Continued}

 \label{symbols}

\begin{tabular}{@{}|ll|c|cc|cc|cc|cc|c|}

  \bf Galaxy & \bf Region & \bf Ref. & \bf [OII] & \bf $\Delta$ &\bf [OIII] &\bf

$\Delta$ & \bf [SII] & \bf $\Delta$ & \bf [SIII] & \bf $\Delta$ & \bf c(H$\beta$)  \\ 

        &   &   & 3727+ 3729 &  & 4959 + 5007 & & 6716+6731 & & 9069+9532 & & \\                   

 \hline

NGC2403 & VS35      &   5   &  0.39   &  0.06   &   0.26  &   0.04  &  -0.32  &   0.01  &  -0.10  &   0.03  &   1.06  \\

    --    & VS24      &   5   &  0.38   &  0.04   &   0.24  &   0.04  &  -0.46  &   0.01  &  -0.04  &   0.04  &  -0.07  \\

    --    & VS38      &   5   &  0.28   &  0.05   &   0.13  &   0.04  &  -0.52  &   0.01  &  -0.07  &   0.05  &   0.28  \\

    --    & VS44      &   5   &  0.45   &  0.14   &   0.30  &   0.04  &  -0.41  &   0.01  &  -0.15  &   0.05  &   0.40  \\

    --    & VS51      &   5   &  0.36   &  0.13   &   0.37  &   0.04  &  -0.47  &   0.01  &  -0.10  &   0.07  &   0.52  \\

    --    & VS3       &   5   &  0.35   &  0.09   &   0.33  &   0.04  &  -0.49  &   0.01  &  -0.11  &   0.07  &   0.60  \\

    --    & VS49      &   5   &  0.34   &  0.05   &   0.51  &   0.04  &  -0.39  &   0.01  &  -0.13  &   0.06  &  -0.10  \\

NGC2541  & 6        &   21   &  0.35   &  0.09   &   0.66  &   0.19  &  -0.50  &   0.19  &  -1.15  &   0.36  &   0.38  \\

      --   & 16        &   21   &  0.28   &  0.07   &   0.74  &   0.09  &  -0.55  &   0.07  &  -0.21  &   0.16  &   0.44  \\

     --   & 17        &   21   &  0.40   &  0.04   &   0.63  &   0.09  &  -0.47  &   0.09  &  -0.38  &   0.15  &   0.36  \\

     --   & 19        &   21   &  0.48   &  0.04   &   0.46  &   0.09  &  -0.21  &   0.07  &  -0.21  &   0.16  &   0.42  \\

UGC4483   & --     &   18  & -0.08   &  0.03   &   0.61  &   0.02  &  -0.93  & 0.01  &  -0.54  &   0.07  &   0.10  \\

NGC2903 & 4        &   21   &  0.24   &  0.04   &  -0.01  &   0.09  &  -0.30  &   0.06  &  -0.21  &   0.14  &   0.28  \\

     --    & 5        &   21   &  0.00   &  0.05   &  -0.51  &   0.12  &  -0.26  &   0.07  &  -0.18  &   0.17  &   0.76  \\

     --    & 6        &   21   & -0.36   &  0.08   &  -0.87  &   0.16  &  -0.41  &   0.07  &  -0.62  &   0.15  &   0.50  \\

     --   & 18        &   21   & -0.20   &  0.08   &  -0.92  &   0.24  &  -0.27  &   0.12  &  -0.56  &   0.24  &   0.95  \\

NGC2903  & R1        &  15   & -0.31   &  0.02   &  -0.76  &   0.03  &  -0.28  &   0.02  &  -0.56  &   0.10  &   0.82  \\

    --    & R2        &  15   & -0.37   &  0.03   &  -0.97  &   0.05  &  -0.28  &   0.02  &  -0.76  &   0.08  &   0.74  \\

    --    & R3        &  15   & -0.17   &  0.04   &  -0.41  &   0.04  &  -0.14  &   0.02  &  -0.16  &   0.13  &   0.46  \\

    --    & R4        &  15   & -0.24   &  0.04   &  -0.76  &   0.03  &  -0.23  &   0.02  &  -0.69  &   0.07  &   0.92  \\

    --    & R6        &  15   & -0.46   &  0.04   &  -1.03  &   0.06  &  -0.28  &   0.02  &  -0.69  &   0.07  &   0.79  \\

    --    & R7        &  15   & -0.41   &  0.03   &  -0.83  &   0.04  &  -0.24  &   0.01  &  -0.56  &   0.10  &   0.99  \\

    --    & R8        &  15   & -0.32   &  0.02   &  -0.76  &   0.03  &  -0.25  &   0.02  &  -0.69  &   0.07  &   0.96  \\

IZw18    &  -- &   4   & -0.59   &  0.06   &   0.46  &   0.01  &  -1.18  &   0.09  &  -0.85  &   0.18  &   0.04  \\

IZw18     &   SE     &  17   & -0.33   &  0.03   &   0.39  &   0.09  &  -1.15  &   0.34  &  -1.02  &   0.16  &   0.20  \\

          &   NW     &  17   & -0.59   &  0.02   &   0.44  &   0.09  &  -1.45  &   0.34  &  -1.09  &   0.16  &   0.10  \\

NGC3310  & Nuc      &   14   &  0.61   &  0.01   &   0.25  &   0.02  &   0.13  &   0.01  &  -0.01  &   0.01  &   0.63  \\

    --   & A          &   14   &  0.41   &  0.01   &   0.48  &   0.01  &  -0.31  &   0.01  &  -0.37  &   0.01  &   0.29  \\

    --   & B          &   14   &  0.49   &  0.01   &   0.38  &   0.01  &  -0.16  &   0.01  &  -0.22  &   0.01  &   0.20  \\

    --   & C          &   14   &  0.50   &  0.01   &   0.51  &   0.01  &  -0.25  &   0.01  &  -0.09  &   0.01  &   0.34  \\

    --   & E          &   14   &  0.50   &  0.01   &   0.44  &   0.01  &  -0.21  &   0.02  &  -0.12  &   0.01  &   0.26  \\

    --   & L          &   14   &  0.67   &  0.01   &   0.36  &   0.02  &  -0.07  &   0.02  &  -0.17  &   0.02  &   0.53  \\

    --   & M          &   14   &  0.40   &  0.01   &   0.53  &   0.03  &  -0.22  &   0.12  &  -0.11  &   0.10  &   0.14  \\ 

NGC3351 & R1        &  15   & -0.37   &  0.03   &  -1.03  &   0.06  &  -0.35  &   0.02  &  -0.38  &   0.03  &   0.81  \\

    --    & R2        &  15   & -0.52   &  0.04   &  -0.92  &   0.05  &  -0.33  &   0.02  &  -0.56  &   0.10  &   0.24  \\

    --    & R3        &  15   & -0.52   &  0.03   &  -1.10  &   0.07  &  -0.37  &   0.03  &  -0.56  &   0.06  &   0.64  \\

    --    & R4        &  15   & -0.60   &  0.05   &  -1.18  &   0.08  &  -0.35  &   0.04  &  -0.35  &   0.06  &   0.24  \\

    --    & R5        &  15   & -0.26   &  0.03   &  -0.53  &   0.04  &  -0.21  &   0.04  &  -0.42  &   0.04  &   0.35  \\

    --    & R7        &  15   & -0.60   &  0.11   &  -1.10  &   0.07  &  -0.41  &   0.07  &  -0.76  &   0.15  &   0.28  \\

    --    & R8        &  15   & -0.60   &  0.06   &  -1.03  &   0.06  &  -0.37  &   0.05  &  -0.86  &   0.18  &   0.36  \\

Mkn 36    & --  &   4   &  0.10   &  0.04   &   0.85  &  0.04   &  -0.60  &   0.06  &  -0.54  &   0.17  &   0.12  \\

NGC3504 & R2        &  15   & -0.15   &  0.02   &  -0.80  &   0.03  &  -0.18  &   0.03  &  -0.29  &   0.03  &   0.28  \\

    --    & R3        &  15   & -0.12   &  0.02   &  -0.62  &   0.02  &  -0.19  &   0.02  &  -0.35  &   0.03  &   0.59  \\

    --    & R4        &  15   & -0.15   &  0.04   &  -0.55  &   0.02  &  -0.12  &   0.03  &  -0.38  &   0.03  &   0.49  \\

NGC3521 & 5        &   21   &  0.07   &  0.09   &  -0.96  &   0.16  &  -0.37  &   0.13  &  -0.69  &   0.30  &   0.98  \\

     --    & 7        &   21   &  0.41   &  0.05   &  -0.46  &   0.20  &  -0.20  &   0.12  &   0.08  &   0.23  &   1.24  \\

     --    & 8        &   21   &  0.61   &  0.05   &   0.62  &   0.12  &  -0.35  &   0.12  &   0.00  &   0.20  &   0.93  \\

     --    & 9        &   21   &  0.54   &  0.07   &   0.17  &   0.16  &  -0.23  &   0.14  &   0.24  &   0.22  &   0.75  \\

NGC3621 & 1        &   21   &  0.32   &  0.03   &  -0.02  &   0.08  &  -0.23  &   0.05  &  -0.51  &   0.16  &   0.31  \\

     --    & 2        &   21   &  0.22   &  0.06   &  -0.10  &   0.11  &  -0.27  &   0.08  &  -0.69  &   0.18  &   0.13  \\

     --    & 3        &   21   &  0.50   &  0.04   &   0.24  &   0.10  &  -0.26  &   0.07  &  -1.16  &   0.18  &   0.84  \\

NGC4253 & A        &  6   &  0.20   &  0.07   &   0.71  &   0.03  &  -0.08  &   0.04  &  -0.03  &   0.06  &   0.51  \\

     --    & B        &  6   &  0.43   &  0.05   &   0.65  &   0.03  &  -0.10  &   0.03  &  -0.05  &   0.06  &   0.45  \\

NGC4254 & 142        &  10   & -0.49   &  0.10   &  -0.97  &   0.14  &  -0.55  &   0.03  &  -0.46  &   0.04  &   0.77  \\

     --   & 78        &  10   & -0.06   &  0.06   &  -1.27  &   0.30  &  -0.40  &   0.03  &  -0.51  &   0.05  &   0.82  \\

     --  & 173        &  10   & -0.20   &  0.05   &  -1.03  &   0.10  &  -0.46  &   0.02  &  -0.42  &   0.04  &   0.38  \\

     --  & 185        &  10   &  0.16   &  0.09   &  -0.32  &   0.05  &  -0.32  &   0.02  &  -0.62  &   0.06  &   1.20  \\

     --  & 184        &  10   &  0.28   &  0.05   &   0.06  &   0.02  &  -0.19  &   0.03  &  -0.42  &   0.04  &   0.32  \\

     --   & 84        &  10   &  0.49   &  0.05   &  -0.19  &   0.05  &  -0.19  &   0.04  &  -0.12  &   0.06  &   0.58  \\

     --   & 22        &  10   &  0.51   &  0.03   &   0.03  &   0.02  &  -0.22  &   0.02  &  -0.12  &   0.03  &   0.59  \\

\hline

  \end{tabular}

 \end{minipage}

 \end{table*}

\begin{table*}

 \begin{minipage}{180mm}

 \setcounter{table}{0}%

\caption{Continued}

 \label{symbols}

\begin{tabular}{@{}|ll|c|cc|cc|cc|cc|c|}

  \bf Galaxy & \bf Region & \bf Ref. & \bf [OII] & \bf $\Delta$ &\bf [OIII] &\bf

$\Delta$ & \bf [SII] & \bf $\Delta$ & \bf [SIII] & \bf $\Delta$ & \bf c(H$\beta$)  \\ 

        &   &   & 3727+ 3729 &  & 4959 + 5007 & & 6716+6731 & & 9069+9532 & & \\                   

 \hline

     --   & 12        &  10   &  0.52   &  0.06   &  -0.25  &   0.07  &  -0.12  &   0.05  &  -0.08  &   0.07  &   0.29  \\

NGC4303 & 95        &  11   &  0.09   &  0.12   &  -0.76  &   0.19  &  -0.29  &   0.11  &  -0.29  &   0.22  &   0.32  \\

     --  & 115        &  11   &  0.19   &  0.03   &  -0.80  &   0.15  &  -0.28  &   0.06  &  -0.69  &   0.18  &   0.78  \\ 

     --  & 155        &  11   &  0.11   &  0.04   &  -0.70  &   0.05  &  -0.32  &   0.05  &  -0.46  &   0.08  &   0.43  \\

     --   & 51        &  11   &  0.45   &  0.12   &  -0.51  &   0.18  &  -0.28  &   0.09  &  -0.76  &   0.20  &   0.74  \\

     --   & 76        &  11   &  0.40   &  0.08   &  -0.22  &   0.06  &  -0.29  &   0.09  &  -0.51  &   0.16  &   0.52  \\

     --   & 53        &  11   &  0.26   &  0.12   &  -0.48  &   0.25  &  -0.12  &   0.22  &  -0.69  &   0.18  &   0.38  \\

     --  & 103        &  11   &  0.24   &  0.13   &  -0.60  &   0.13  &  -0.26  &   0.12  &  -0.51  &   0.19  &   0.55  \\

     --  & 124        &  11   &  0.50   &  0.17   &  -0.48  &   0.17  &  -0.12  &   0.11  &  -0.42  &   0.13  &   0.48  \\

     --  & 148        &  11   &  0.48   &  0.08   &  -0.12  &   0.08  &  -0.19  &   0.08  &  -0.56  &   0.21  &   0.20  \\

     --  & 278        &  11   &  0.46   &  0.05   &   0.42  &   0.05  &  -0.21  &   0.08  &  -0.69  &   0.15  &   0.13  \\

     --  & 234        &  11   &  0.65   &  0.06   &   0.55  &   0.06  &   0.02  &   0.07  &  -0.12  &   0.21  &   0.00  \\

NGC4528 & 4        &   21   &  0.29   &  0.03   &   0.16  &   0.07  &  -0.62  &   0.05  &   0.12  &   0.18  &   0.37  \\

NGC4559 & 1        &   21   &  0.43   &  0.04   &   0.02  &   0.11  &  -0.13  &   0.09  &  -0.14  &   0.15  &   0.46  \\

     --    & 3        &   21   &  0.43   &  0.04   &   0.52  &   0.10  &  -0.32  &   0.07  &  -0.10  &   0.14  &   0.17  \\

     --    & 4        &   21   &  0.39   &  0.04   &   0.60  &   0.09  &  -0.58  &   0.08  &  -0.86  &   0.18  &   0.17  \\

     --   & 17        &   21   &  0.41   &  0.04   &   0.38  &   0.09  &  -0.28  &   0.07  &   0.12  &   0.16  &   0.49  \\

     --   & 18        &   21   &  0.48   &  0.03   &   0.35  &   0.07  &  -0.43  &   0.06  &   0.07  &   0.15  &   0.31  \\

     --   & 19        &   21   &  0.47   &  0.03   &   0.05  &   0.11  &  -0.28  &   0.08  &   0.01  &   0.17  &   0.45  \\

     --   & 20        &   21   &  0.43   &  0.04   &   0.00   &   0.10  &  -0.35  &   0.08  &   0.04  &   0.16  &   0.53  \\

NGC4861   & --      &   4   &  0.11   &  0.04   &   0.92  &  0.04   &  -0.75  &   0.06  &  -0.32  &   0.07  &   0.09  \\

M51   &  CCM72    &   3   & -0.21   &  0.02   &  -1.18  &   0.15  &  -0.43  &   0.04  &  -0.29  &   0.06  &   0.31  \\

     --   &  CCM24    &   3   &  0.01   &  0.01   &  -0.65  &   0.20  &  -0.38  &   0.03  &  -0.48  &   0.15  &   0.60  \\

     --   &  CCM10    &   3   &  0.21   &  0.01   &  -0.58  &   0.04  &  -0.40  &   0.06  &  -0.48  &   0.04  &   0.30  \\

M101  & H602       &   13   & -0.59   &  0.03   &  -0.96  &   0.10  &  -0.47  &   0.04  &  -0.68  &   0.12  &   0.39  \\

 --    & H493       &   13  &  0.05   &  0.11   &  -0.89  &   0.09  &  -0.32  &   0.03  &  -0.37  &   0.09  &   0.50  \\

 --    & H507       &   13  & -0.19   &  0.03   &  -1.00  &   0.04  &  -0.31  &   0.02  &  -0.41  &   0.09  &   0.46  \\

 --    & H972       &   13  &  0.14   &  0.02   &  -0.40  &   0.03  &  -0.47  &   0.03  &  -0.14  &   0.08  &   0.30  \\

 --    & H942       &   13  &  0.17   &  0.04   &  -0.41  &   0.07  &  -0.25  &   0.04  &  -0.30  &   0.10  &   0.52  \\

 --    & H974       &   13  &  0.21   &  0.03   &  -0.47  &   0.10  &  -0.34  &   0.05  &  -0.66  &   0.16  &   0.33  \\

 --    & H959       &   13  &  0.48   &  0.09   &  -0.42  &   0.05  &  -0.21  &   0.02  &  -0.28  &   0.09  &   0.52  \\

 --    & H1013      &   13  &  0.26   &  0.02   &   0.12  &   0.02  &  -0.64  &   0.02  &   0.06  &   0.08  &   0.30  \\

 --    & H399       &   13  &  0.19   &  0.05   &  -0.18  &   0.03  &  -0.46  &   0.02  &  -0.03  &   0.08  &   0.32  \\

 --    & H1044      &   13  &  0.28   &  0.03   &  -0.12  &   0.03  &  -0.44  &   0.02  &  -0.06  &   0.08  &   0.28  \\

 --    & H953       &   13  &  0.27   &  0.06   &  -0.25  &   0.04  &  -0.37  &   0.02  &  -0.33  &   0.08  &   0.24  \\

 --    & H949       &   13  &  0.23   &  0.05   &  -0.43  &   0.03  &  -0.30  &   0.02  &  -0.17  &   0.08  &   0.39  \\

 --    & H1052      &   13  &  0.32   &  0.04   &   0.50  &   0.02  &  -0.64  &   0.02  &   0.12  &   0.08  &   0.29  \\

 --    & H203       &   13  &  0.36   &  0.02   &  -0.27  &   0.02  &  -0.26  &   0.02  &  -0.17  &   0.08  &   0.22  \\

 --    & H1086      &   13  &  0.43   &  0.05   &   0.37  &   0.03  &  -0.46  &   0.01  &  -0.09  &   0.08  &   0.32  \\

 --    & H875       &   13  &  0.53   &  0.05   &   0.09  &   0.03  &  -0.29  &   0.03  &  -0.21  &   0.08  &   0.43  \\

 --    & H1098      &   13  &  0.49   &  0.04   &   0.18  &   0.02  &  -0.40  &   0.02  &   0.04  &   0.08  &   0.74  \\

 --    & H237       &   13  &  0.11   &  0.02   &   0.19  &   0.02  &  -0.49  &   0.03  &  -0.25  &   0.08  &   0.20  \\

 --    & H1026      &   13  &  0.37   &  0.02   &   0.57  &   0.02  &  -0.57  &   0.02  &  -0.07  &   0.08  &   0.11  \\

 --    & H178       &   13  &  0.65   &  0.04   &   0.26  &   0.04  &  -0.43  &   0.02  &  -0.14  &   0.08  &   0.59  \\

 --    & H1159      &   13  &  0.40   &  0.03   &   0.61  &   0.02  &  -0.64  &   0.02  &  -0.04  &   0.08  &   0.25  \\

   --  & H1170      &   13  &  0.38   &  0.02   &   0.63  &   0.02  &  -0.54  &   0.03  &  -0.03  &   0.08  &   0.39  \\

   --  & H1176      &   13  &  0.32   &  0.02   &   0.62  &   0.02  &  -0.70  &   0.02  &   0.00  &   0.08  &   0.39  \\

   --  & H140       &   13  &  0.45   &  0.02   &   0.20  &   0.02  &  -0.11  &   0.02  &  -0.19  &   0.08  &   0.13  \\

   --  & H1191      &   13  &  0.34   &  0.03   &   0.67  &   0.03  &  -0.60  &   0.03  &  -0.13  &   0.09  &   0.35  \\

   --  & H119       &   13  &  0.49   &  0.09   &   0.35  &   0.03  &  -0.12  &   0.03  &  -0.52  &   0.10  &   0.33  \\

   --  & H132       &   13  &  0.48   &  0.02   &   0.46  &   0.02  &  -0.34  &   0.02  &  -0.48  &   0.08  &   0.28  \\

   --  & H143       &   13  &  0.45   &  0.02   &   0.54  &   0.02  &  -0.36  &   0.02  &  -0.08  &   0.08  &   0.39  \\

   --  & H149       &   13  &  0.45   &  0.02   &  -0.39  &   0.02  &  -0.42  &   0.01  &  -0.06  &   0.08  &   0.45  \\

   --  & H128       &   13 &  0.26   &  0.02   &   0.72  &   0.02  &  -0.70  &   0.02  &  -0.01  &   0.08  &   0.27  \\

   --  & H67        &   13 &  0.37   &  0.02   &   0.68  &   0.02  &  -0.68  &   0.04  &  -0.13  &   0.08  &   0.07  \\

   --  & H1216      &   13 &  0.29   &  0.03   &   0.79  &   0.02  &  -0.85  &         0.03  &  -0.07  &   0.08  &   0.17  \\

   --  & H681       &   13 &  0.39   &  0.03   &   0.61  &   0.03  &  -0.57  &   0.05  &  -0.44  &   0.10  &   0.22  \\

   --  & NGC5471    &   16  &  0.06   &  0.04   &   0.89  &   0.04  &  -0.61  &   0.05  &  -0.20  &   0.05  &   0.18  \\

   --  & NGC5471    &    4  &  0.12   &  0.01   &   0.96  &   0.01  &  -0.70  & 0.05  &  -0.32  &   0.07  &   0.15  \\            

   --  & NGC5471A     &   13  &  0.11   &  0.03   &   0.97  &   0.02  &  -0.82  &   0.03  &  -0.30  &   0.08  &   0.22  \\

\hline

  \end{tabular}

 \end{minipage}

 \end{table*}

\begin{table*}

 \begin{minipage}{180mm}

 \setcounter{table}{0}%

\caption{Continued}

 \label{symbols}

\begin{tabular}{@{}|ll|c|cc|cc|cc|cc|c|}

  \bf Galaxy & \bf Region & \bf Ref. & \bf [OII] & \bf $\Delta$ &\bf [OIII] &\bf

$\Delta$ & \bf [SII] & \bf $\Delta$ & \bf [SIII] & \bf $\Delta$ & \bf c(H$\beta$)  \\ 

        &   &   & 3727+ 3729 &  & 4959 + 5007 & & 6716+6731 & & 9069+9532 & & \\                   

 \hline

   --  & NGC5471A     &    2  &  0.17   &  0.03   &   0.81  &   0.02  &  -0.73  &   0.03  &  -0.35  &   0.03  &   0.25  \\ 

   --  & NGC5471B     &   13  &  0.45   &  0.03   &   0.75  &   0.02  &  -0.27  &   0.02  &  -0.30  &   0.08  &   0.23  \\

   --  & NGC5471C     &   13  &  0.36   &  0.03   &   0.80  &   0.02  &  -0.60  &   0.02  &  -0.20  &   0.08  &   0.19  \\

   --  & NGC5471D     &   13  &  0.18   &  0.03   &   0.91  &   0.02  &  -0.80  &   0.03  &  -0.13  &   0.08  &   0.09  \\

   --  & NGC5471E     &   13  &  0.10   &  0.03   &   0.91  &   0.02  &  -0.92  &   0.04  &  -0.14  &   0.08  &   0.24  \\

   --  & S5         &   13  &  0.28   &  0.02   &  -0.48  &   0.03  &  -0.29  &   0.02  &  -0.10  &   0.08  &   0.30  \\

   --  & S5         &   16  &  0.26   &  0.05   &  -0.49  &   0.06  &  -0.19  &   0.05  &   0.07  &   0.10  &   0.22  \\

   --  & S5         &    2  &  0.32   &  0.04   &  -0.47  &   0.03  &  -0.17  & 0.05  &   0.30  &   0.07  &   0.50  \\

   --  & NGC5455    &   13  &  0.32   &  0.02   &   0.68  &   0.02  &  -0.52  &   0.01  &  -0.08  &   0.08  &   0.12  \\

   --  & NGC5455    &   16  &  0.41   &  0.04   &   0.68  &   0.04  &  -0.43  &   0.05  &   0.00  &   0.05  &   0.15  \\

   --  & NCG5461    &   13  &  0.22   &  0.02   &   0.64  &   0.02  &  -0.68  &   0.02  &   0.08  &   0.08  &   0.36  \\

   --  & NGC5461A   &    2  &  0.27   &  0.02   &   0.64  &   0.02  &  -0.47  & 0.02  &   0.00  &   0.01  &   0.60  \\

NGC5953 & A        &   7   & -0.10   &  0.10   &  -0.52  &   0.04  &  -0.20  &   0.02  &  -0.46  &   0.04  &   0.00  \\

     -- & B        &   7   & -0.05   &  0.09   &  -0.44  &   0.06  &  -0.19  &   0.03  &  -0.49  &   0.05  &   0.00  \\

     -- & C        &   7   &  0.20   &  0.07   &   0.36  &   0.02  &  -0.02  &   0.02  &  -0.44  &   0.03  &   0.00  \\

IZw123    & --  &   4   &  0.10   &  0.04   &   0.96  &  0.04   &  -0.64  &   0.06  &  -0.45  &   0.15  &   0.38  \\ 

NGC7714 & A        &   9   &  0.39   &  0.01   &   0.25  &   0.01  &  -0.13  &   0.01  &  -0.21  &   0.02  &   0.33  \\

     -- & N110     &   9   &  0.26   &  0.02   &   0.27  &   0.01  &  -0.24  &   0.03  &  -0.02  &   0.05  &   0.30  \\

     -- & B        &   9   &  0.38   &  0.01   &   0.68  &   0.01  &  -0.28  &   0.02  &  -0.31  &   0.02  &   0.37  \\

     -- & C        &   9   &  0.45   &  0.04   &   0.54  &   0.01  &  -0.24  &   0.05  &  -0.38  &   0.09  &   0.44  \\

     -- & N216     &   9   &  0.24   &  0.02   &   0.31  &   0.01  &  -0.23  &   0.03  &   0.03  &   0.05  &   0.30  \\

\hline

  \end{tabular}

{\bf References to the table}

1:  Dennefeld \& Stasi\'nska  (1983); 2: D\'\i az {\it et al.} (1990); 

3:  D\'\i az {\it et al.} (1991); 4: Garnett (1989); 5: Garnett {\it et al.}  (1997); 6: Gonz\'alez-Delgado \& P\'erez (1996a); 7: Gonz\'alez-Delgado \& P\'erez, (1996b); 8: Gonz\'alez-Delgado {\it et al.} (1994);

9: Gonz\'alez-Delgado 

{\it et al.} (1995); 10: Henry {\it et al.} (1992); 11: Henry {\it et al.}

(1994); 12: Henry {\it et al.} (1996);

13: Kennicutt \& Garnett (1996); 14: Pastoriza {\it et al.}

(1993); 15: P\'erez-Olea (1996); 16: Shields \& Searle (1978);

17: Skillman \& Kennicutt (1993);

18: Skillman {\it et al.} (1993); 19: V\'\i lchez \& Esteban (1996); 

20: V\'\i lchez {\it et al.} (1988); 21: Zaritsky, Kennicutt \& Huchra (1994).

 \end{minipage}

 \end{table*}

Only 65 of these objects have electron temperatures derived directly. Table 
2 lists for these objects their electron temperature, in units of 10$^4$ K, and 
their derived oxygen abundance expressed as $12+log \frac{O}{H}$.
The electron temperature has been derived from the measured intensity of the [OIII] \lam\ 4363 line except for the objects marked with an asterisk for which 
a measurement of the [OII] \lam\ 7327 line has been used.  The total 
oxygen abundance has been calculated as 
$log\frac{O}{H}=log\left(\frac{O^+}{H^+}+\frac{O^{++}}{H^+}\right)$. The errors in the abundances have been derived from those quoted for the emission lines 
whenever possible. For the HII regions in the Magellanic Clouds (Dennefeld \& Stasi\'ska 1983) we have adopted an abundance error of $\pm$ 0.10.
 
 The oxygen abundance ranges from 2\% solar for IZw18 to 70 \% solar for region 
VS24 in NGC~2403. 

We have included in the table data for three regions of metallicity close to 
solar for which detailed modelling has been made: CC93 in M33 (V\'\i lchez 
{\it et al.} 1988), S5 in M101 (Shields \& Searle 1978; D\'\i az {\it et al.}
1990) and the nucleus of NGC~3310 (Pastoriza {\it et al.} 1993).

%%%%%%%%%%%%%%%%%%%%%%%%%%%%%%%%%%%%%%%%%%%%%%%%%%%%%%%%%%%%%%%%%%%%%%%%%%%%%%%%

%

%             Tabla 2 Datos para la calibracion

%

%%%%%%%%%%%%%%%%%%%%%%%%%%%%%%%%%%%%%%%%%%%%%%%%%%%%%%%%%%%%%%%%%%%%%%%%%%%%%%%%

\begin{table*}

 \begin{minipage}{130mm}

 \caption{Oxygen abundance and abundance parameters for the sample objects}

 \label{symbols}

\begin{tabular}{|c|c|c|c|c|c|c|}

\hline \hline

Galaxy  & Region & Ref & $log O_{23}$ & $log S_{23}$ & $12+log(O/H)$ & $t_{e} (10^4K)$  \\

\hline

MWG     &N2467   & 1       &0.75 $\pm$ 0.04      &-0.03          &8.01 $\pm$ 0.10      &1.18     \\

--      &ETACAR  & 1       &0.66 $\pm$ 0.04      & 0.18          &8.37$\pm$ 0.10       &0.89     \\

--      &M17     & 1       &0.75 $\pm$ 0.04      & 0.22          &8.43$\pm$ 0.10       &0.88     \\ 

--      &M16     & 1       &0.55 $\pm$ 0.04      &-0.02          &8.55$\pm$ 0.10       &1.81     \\ 

--      &M20     & 1       &0.66 $\pm$ 0.04      & 0.20          &8.50$\pm$ 0.10       &2.73     \\

--      &NGC3576 & 1       &0.76 $\pm$ 0.04      & 0.46          &8.42$\pm$ 0.10       &0.91     \\ 

--      &Orion1  & 1       &0.69 $\pm$ 0.04      & 0.04          &8.54$\pm$ 0.10       &0.84     \\ 

--      &Orion2  & 1       &0.76 $\pm$ 0.04      & 0.26          &8.50$\pm$ 0.10       &0.87     \\

--      &S127 $^{*}$   & 19      &0.62 $\pm$ 0.04      &-0.12 $\pm$ 0.06     &8.06$\pm$ 0.26       &0.80     \\ 

--      &S128 $^{*}$   & 19      &0.78 $\pm$ 0.02      &-0.09 $\pm$ 0.06     &8.04$\pm$ 0.24       &0.90     \\ 

IC10    &\# 2    & 4       &0.93 $\pm$ 0.01      & 0.09 $\pm$ 0.06          &8.26 $\pm$ 0.10      &1.08     \\

SMC     &N80     & 1       &0.84 $\pm$ 0.04      &-0.25          &7.85$\pm$ 0.10      &1.39     \\ 

--      &N83     & 1       &0.82 $\pm$ 0.04      &-0.09          &8.01$\pm$ 0.10       &1.21     \\ 

--      &N13     & 1       &0.92 $\pm$ 0.04      &-0.08               &8.02 $\pm$ 0.10      &1.30     \\ 

--      &N32     & 1       &0.72 $\pm$ 0.04      &-0.18             &8.16$\pm$ 0.10       &2.13     \\ 

--      &N81     & 1       &0.92 $\pm$ 0.04      &-0.07               &8.05 $\pm$ 0.10      &1.26     \\ 

--      &N66     & 1       &0.90 $\pm$ 0.04      &-0.31               &7.96 $\pm$ 0.10      &1.31     \\

M33     &N604    & 4       &0.70 $\pm$ 0.01      & 0.04 $\pm$ 0.02      &8.51$\pm$ 0.03       &0.82     \\

--      &N604    & 20      &0.70 $\pm$ 0.01      & 0.00 $\pm$ 0.02      &8.51$\pm$ 0.03       &0.77     \\ 

--      &N595    & 4       &0.61 $\pm$ 0.01      & 0.09 $\pm$ 0.02      &8.44$\pm$ 0.09       &0.90     \\ 

--      &N595    & 20      &0.61 $\pm$ 0.01      & 0.04 $\pm$ 0.02      &8.44$\pm$ 0.09       &0.80     \\ 

--      &N588    & 4       &0.89 $\pm$ 0.01      & 0.09 $\pm$ 0.03          &8.33 $\pm$ 0.06      &0.94     \\

--      &N588    & 20      &0.89 $\pm$ 0.01      & 0.03 $\pm$ 0.03          &8.33 $\pm$ 0.06      &1.01     \\

--      &MA2     & 20      &0.61 $\pm$ 0.01      & 0.09 $\pm$ 0.02          &8.44 $\pm$ 0.15      &0.80     \\ 

--      &IC142   & 20      &0.52 $\pm$ 0.01      & 0.19 $\pm$ 0.03      &8.69$\pm$ 0.16       &0.69     \\ 

--      &IC131   & 4       &0.87 $\pm$ 0.01      & 0.06 $\pm$ 0.08          &8.41 $\pm$ 0.06      &0.95     \\

--      &CC93    & 20      &0.28 $\pm$ 0.15      & 0.28 $\pm$ 0.04          &9.02 $\pm$ 0.16      &0.55     \\

Mkn600  &  --    & 4       &1.02 $\pm$ 0.04      &-0.34 $\pm$ 0.12      &7.86$\pm$ 0.10       &1.64     \\

LMC    &N59A    & 1       &0.96 $\pm$ 0.04      &-0.01          &8.40$\pm$ 0.10       &1.01     \\ 

--      &N44B    & 1       &1.09 $\pm$ 0.04      & 0.09          &8.39$\pm$ 0.10       &1.11     \\ 

--      &N55A    & 1       &0.76 $\pm$ 0.04      & 0.04          &8.27$\pm$ 0.10       &0.98     \\ 

--      &N113D   & 1       &0.90 $\pm$ 0.04      & 0.15          &8.68$\pm$ 0.10       &0.87     \\ 

--      &N127A   & 1       &0.88 $\pm$ 0.04      & 0.13               &8.50 $\pm$ 0.10      &0.95     \\ 

--      &N159A   & 1       &0.87 $\pm$ 0.04      & 0.02          &8.23$\pm$ 0.10       &1.07     \\ 

--      &N214C   & 1       &0.86 $\pm$ 0.04      & 0.12          &8.29$\pm$ 0.10       &1.04     \\

--      &N4A     & 1       &0.84 $\pm$ 0.04      & 0.02          &8.43$\pm$ 0.10       &0.93     \\

--      &N79E    & 1       &0.65 $\pm$ 0.04      & 0.00          &8.31$\pm$ 0.10       &0.92     \\ 

--      &N191A   & 1       &0.92 $\pm$ 0.04      & 0.14          &8.53$\pm$ 0.10       &1.09     \\ 

IIZw40  &  --    & 4       &1.04 $\pm$ 0.04      &-0.22 $\pm$ 0.07       &8.10$\pm$ 0.10       &1.33     \\ 

IIZw40  &  --    & 2       &1.01 $\pm$ 0.04      &-0.15 $\pm$ 0.07       &8.14$\pm$ 0.02       &1.28     \\

N2366   &N2363    & 4       &1.00  $\pm$ 0.02     &-0.39 $\pm$ 0.12         &7.92$\pm$ 0.04       &1.48     \\ 

--      &N2363~A1 & 8       &0.86  $\pm$ 0.08     &-0.15 $\pm$ 0.15      &7.74$\pm$ 0.02       &1.62     \\ 

--      &N2363~A2 & 8       &1.02  $\pm$ 0.01     &-0.42 $\pm$ 0.02      &7.87$\pm$ 0.02       &1.61     \\ 

--      &N2363~A3 & 8       &0.95  $\pm$ 0.08     &-0.18 $\pm$ 0.20      &7.82$\pm$ 0.02       &1.59     \\

N2403   &VS35     & 5       &0.63  $\pm$ 0.05     & 0.10 $\pm$ 0.02      &8.41$\pm$ 0.14       &0.84     \\ 

--      &VS24     & 5       &0.62  $\pm$ 0.04     & 0.10 $\pm$ 0.03          &8.76 $\pm$ 0.09      &0.68 \\ 

--      &VS38     & 5       &0.51  $\pm$ 0.05     & 0.06 $\pm$ 0.04      &8.47$\pm$ 0.10       &0.76     \\ 

--      &VS44     & 5       &0.68  $\pm$ 0.10     & 0.04 $\pm$ 0.04      &8.49$\pm$ 0.13       &0.87     \\ 

--      &VS51     & 5       &0.67  $\pm$ 0.08     & 0.05 $\pm$ 0.05      &8.53$\pm$ 0.15       &0.80     \\ 

--      &VS3      & 5       &0.64  $\pm$ 0.07     & 0.04 $\pm$ 0.05      &8.41$\pm$ 0.09       &0.87     \\ 

--      &VS49     & 5       &0.73  $\pm$ 0.04     & 0.06 $\pm$ 0.04      &8.20$\pm$ 0.08       &1.05     \\ 

UGC4483 &  --    & 18      &0.69 $\pm$ 0.02      &-0.39 $\pm$ 0.05      &7.51$\pm$ 0.04       &1.66     \\

IZw18   &  --    & 4       &0.47 $\pm$ 0.01      &-0.67 $\pm$ 0.15      &7.27$\pm$ 0.10       &1.80     \\

IZw18   &SE      & 17      &0.47 $\pm$ 0.01      &-0.78 $\pm$ 0.08      &7.27$\pm$ 0.08       &1.72     \\ 

--      &NW      & 17      &0.48 $\pm$ 0.02      &-0.93 $\pm$ 0.09      &7.17$\pm$ 0.06       &1.96     \\

N3310   &Nuc      & 14      &0.77  $\pm$ 0.01     & 0.37 $\pm$ 0.01      &9.01$\pm$ 0.15      &0.62     \\ 

--      &A        & 14      &0.75  $\pm$ 0.01     &-0.04 $\pm$ 0.01         &8.20 $\pm$ 0.10      &1.04     \\ 

--      &B        & 14      &0.74  $\pm$ 0.01     & 0.11 $\pm$ 0.01      &8.13$\pm$ 0.09       &1.21     \\ 

--      &C        & 14      &0.81  $\pm$ 0.01     & 0.14 $\pm$ 0.01      &8.25$\pm$ 0.04       &0.95     \\

\hline

  \end{tabular}

 \end{minipage}

 \end{table*}

\begin{table*}

 \setcounter{table}{1}

 \begin{minipage}{130mm}

 \caption{Continued}

 \label{symbols}

\begin{tabular}{|c|c|c|c|c|c|c|}

\hline \hline

Object & Region &  Ref. & $log O_{23}$ & $log S_{23}$ & $12+log(O/H)$ &  $t_{e} (10^4K)$  \\

\hline

--      &E        & 14      &0.77  $\pm$ 0.01     & 0.14 $\pm$ 0.01      &8.17$\pm$ 0.08       &0.98     \\ 

--      &L        & 14      &0.84  $\pm$ 0.01     & 0.18 $\pm$ 0.02      &8.47$\pm$ 0.13       &0.70     \\ 

--      &M        & 14      &0.77  $\pm$ 0.02     & 0.14 $\pm$ 0.11      &8.30$\pm$ 0.15       &0.97     \\ 

Mkn36   &  --    & 4       &0.92 $\pm$ 0.04      &-0.27 $\pm$ 0.12           &7.86 $\pm$ 0.10      &1.50 \\   

N4861   &  --    & 4       &1.10 $\pm$ 0.02      &-0.18 $\pm$ 0.07      &8.26$\pm$ 0.10       &1.21     \\  

M101    &N5455   & 16      &0.87 $\pm$ 0.04      & 0.14 $\pm$ 0.05          &8.54 $\pm$ 0.10      &0.89     \\ 

--      &N5471   & 16      &0.95 $\pm$ 0.04       &-0.06 $\pm$ 0.05          &8.13 $\pm$ 0.10      &1.19     \\

--      &N5471   & 2       &1.00 $\pm$ 0.04       &-0.20 $\pm$ 0.05      &8.01$\pm$ 0.10       &1.28     \\

--      &N5471   & 4       &1.00 $\pm$ 0.04       &-0.16 $\pm$ 0.05      &8.13$\pm$ 0.05       &1.30     \\

--      &N5461   & 2       &0.79 $\pm$ 0.04       & 0.13 $\pm$ 0.05          &8.46 $\pm$ 0.08      &0.90   \\ 

--      &S5      & 2       &0.38 $\pm$ 0.05       & 0.48 $\pm$ 0.08      &8.90$\pm$ 0.15       &0.65     \\

--      &S5      & 16      &0.33 $\pm$ 0.06       & 0.26 $\pm$ 0.08          &9.10 $\pm$ 0.20      &0.58   \\ 

IZw123  &  --    & 4       &1.02 $\pm$ 0.02      &-0.23 $\pm$ 0.12      &8.01$\pm$ 0.10       &1.41     \\  

N7714   &A       & 9       &0.62 $\pm$ 0.01       & 0.13 $\pm$ 0.01      &8.47$\pm$ 0.16       &1.26     \\ 

--      &B       & 9       &0.86 $\pm$ 0.01       & 0.01 $\pm$ 0.02          &8.20 $\pm$ 0.09      &1.11   \\ 

--      &C       & 9       &0.80 $\pm$ 0.02       & 0.00 $\pm$ 0.07         &8.27 $\pm$ 0.19      &1.02     \\ 

--      &N110    & 9       &0.57 $\pm$ 0.01       & 0.18 $\pm$ 0.04         &8.53 $\pm$ 0.13      &1.10     \\ 

--      &N216    & 9       &0.58 $\pm$ 0.01       & 0.22 $\pm$ 0.04      &8.46$\pm$ 0.12       &1.01     \\

\hline

  \end{tabular}

{\bf References to the table}

1:  Dennefeld \& Stasi\'nska  (1983); 2: D\'\i az {\it et al.} (1990); 

3:  D\'\i az {\it et al.} (1991); 4: Garnett (1989); 5: Garnett {\it et al.}  (1997); 6: Gonz\'alez-Delgado \& P\'erez (1996a); 7: Gonz\'alez-Delgado \& P\'erez, (1996b); 8: Gonz\'alez-Delgado {\it et al.} (1994);

9: Gonz\'alez-Delgado 

{\it et al.} (1995); 10: Henry {\it et al.} (1992); 11: Henry {\it et al.}

(1994); 12: Henry {\it et al.} (1996);

13: Kennicutt \& Garnett (1996); 14: Pastoriza {\it et al.}

(1993); 15: P\'erez-Olea (1996); 16: Shields \& Searle (1978);

17: Skillman \& Kennicutt (1993);

18: Skillman {\it et al.} (1993); 19: V\'\i lchez \& Esteban (1996); 

20: V\'\i lchez {\it et al.} (1988); 21: Zaritsky, Kennicutt \& Huchra (1994).

\end{minipage}

 \end{table*}

\section{Results and Discussion}

From the line intensities of the sulphur lines presented in Table 1 we have 
calculated the {\it sulphur abundance parameter} $S_{23}$ defined as 
\[ S_{23} = \frac{[SII]\lambda\lambda 6717,6731+ [SIII]\lambda\lambda 9069,9532}{H\beta} \]
 This parameter is analogous to the $R_{23}$ 
parameter defined for the optical oxygen lines and hereafter we will refer to 
the oxygen and sulphur abundance parameters as $O_{23}$ and $S_{23}$ 
respectively. Both parameters are listed in Table 2.

The relation between  $O_{23}$ and the oxygen 
abundance for the objects in Table 2 can be seen in the upper panel of 
Fig 1 (solid dots) together 
with similar data for HII galaxies compiled by D\'\i az (1999) (open circles).
The figure illustrates the problems mentioned in the introduction, the most
important one being the two-valued nature of the relation which makes an 
accurate metallicity calibration virtually impossible, more so for objects 
with $logO_{23} \geq 0.8$. Also notice the position of the nucleus of NGC~3310 
which is probably due to its higher than usual density ($n_H$ = 8000 cm$^{-3}$;
Pastoriza et al. 1993).
 On the contrary, and as expected, the relation between $S_{23}$ and oxygen abundance for the objects of Table 2, shown in the lower panel of the figure, remains single valued up to a metallicity 
close to solar. In both figures the data corresponding to 
different observations of IZw18, NGC~5471 in M101, and S5 in M101 have been joined together to show the importance of internal errors. The solar 
metallicity regions have also been labelled. 

Also, as compared to the case of $O_{23}$, the scatter in the 
relation between S$_{23}$ and oxygen abundance is somewhat reduced.  
 
%%%%%%%%%%%%%%%%%%%%%%%%%%%%%%%%%%%%%%%%%%%%%%%%%%%%%%%%%%%%%%%%%%%%%%%%%%%%%%%%
%
%     Figure 3 Empirical calibration of oxygen abundance through the sulphur 
%              abundance parameter S23. The ressiduals of the fits are shown 
%              in the lower panel of the figure (see text for details) 
%
%
%%%%%%%%%%%%%%%%%%%%%%%%%%%%%%%%%%%%%%%%%%%%%%%%%%%%%%%%%%%%%%%%%%%%%%%%%%%%%%%%
\begin{figure}
%\begin{minipage}{180mm}
\psfig{figure=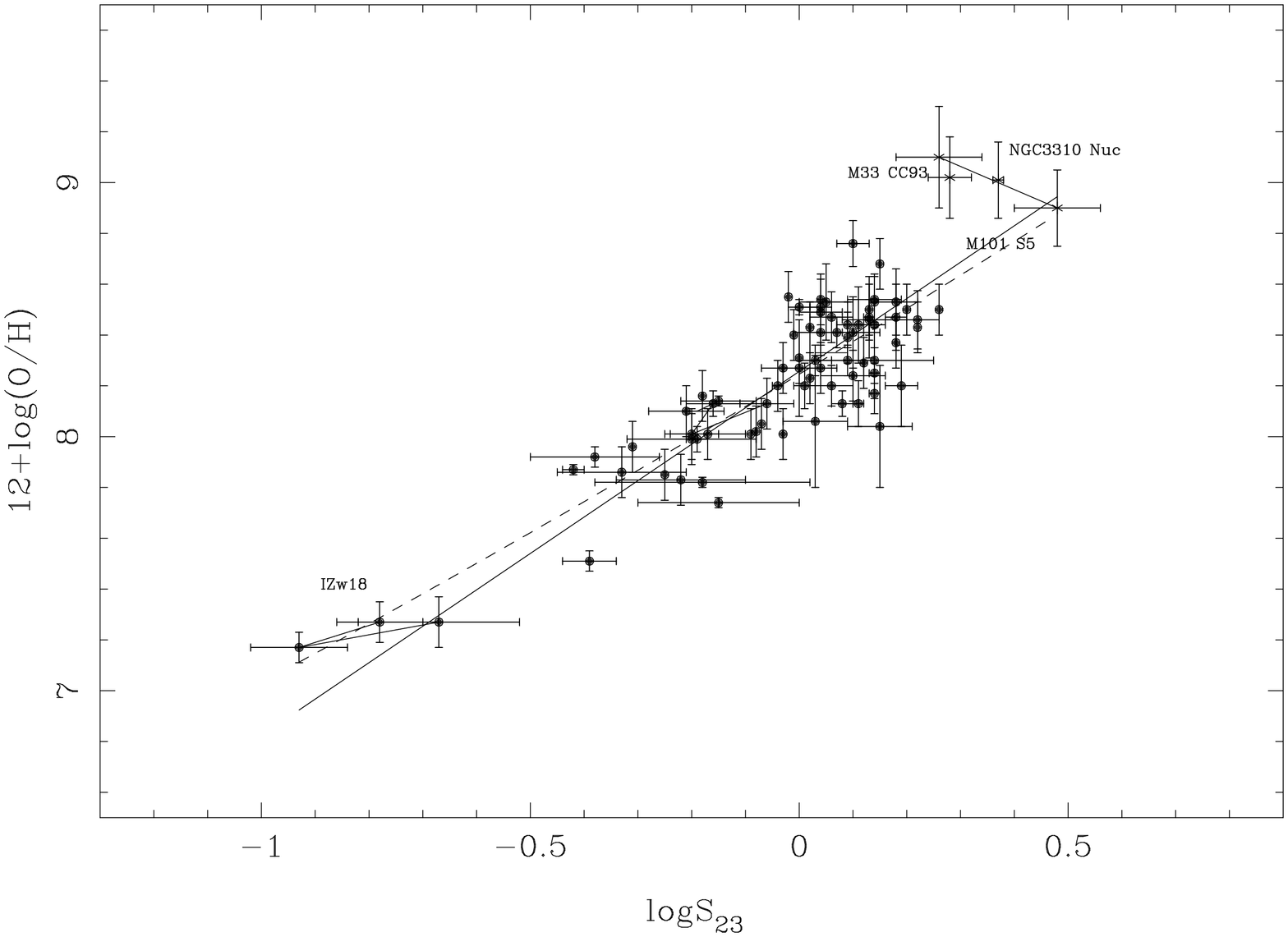,width=7cm,rheight=7.0cm,angle=270}
\psfig{figure=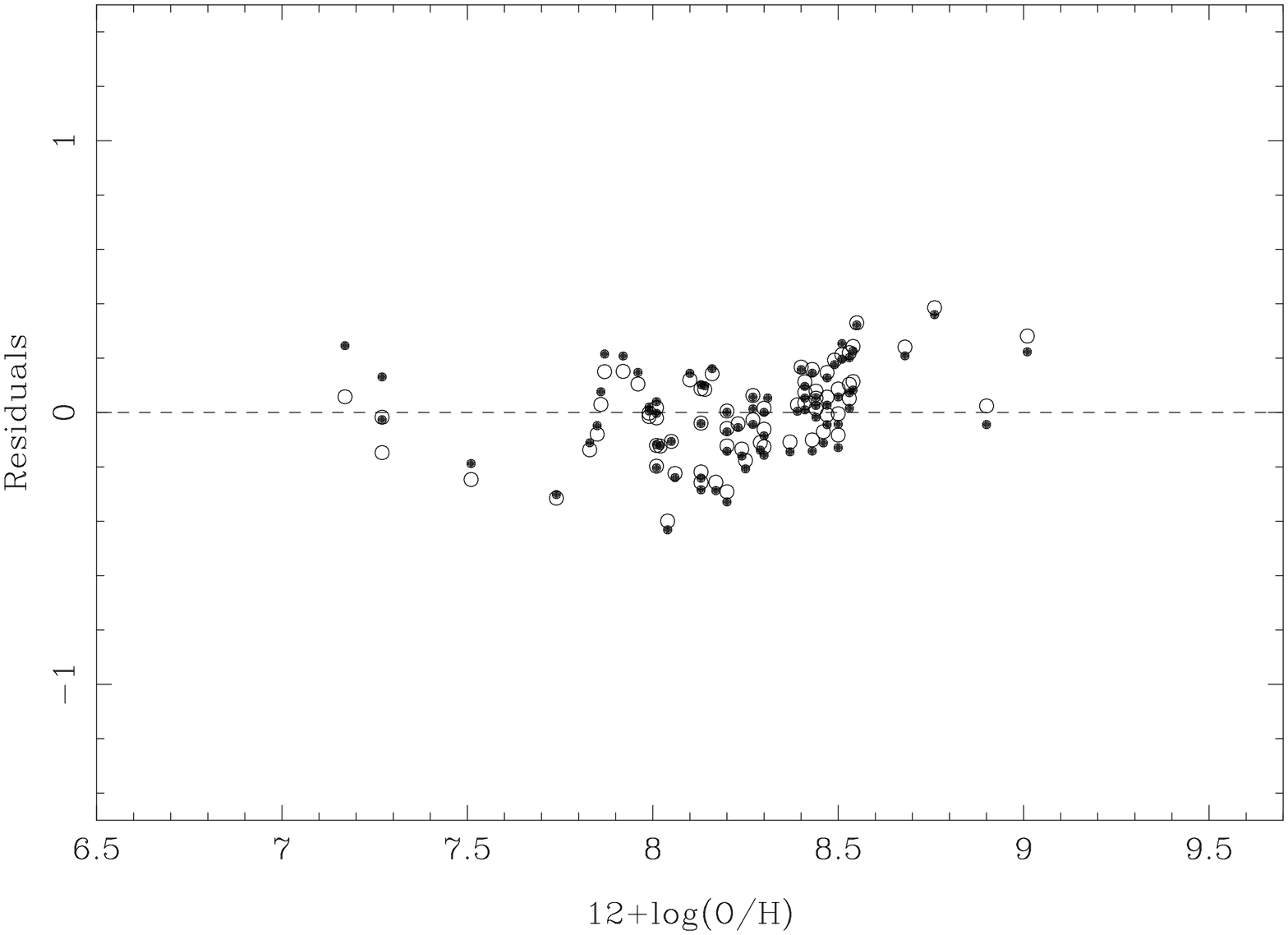,width=7cm,rheight=7.0cm,angle=270}
\caption{Empirical calibration of oxygen abundance through the sulphur 
abundance parameter $S_{23}$. The ressiduals of the fits are shown in the 
lower panel of the figure (see text for details). }
%\end{minipage}
\end{figure}

The dependence of $O_{23}$ on the degree of 
ionization of the nebula is partially responsible for the large 
scatter found in the $O_{23}$ {\it vs} metallicity relation . Figure 2 (upper panel) shows the relation between $logO_{23}$ and $log([OII]/[OIII])$,
which can be taken as a good ionization parameter indicator for ionizing 
temperatures larger than about 35000 K (see D\'\i az 1999), for the objects in Table 1. A positive correlation between ionization parameter and $O_{23}$ is evident for 
objects with $O_{23}$ between 0.2 and 1 (higher excitation objects are at the 
left in the plot).  On the contrary,
no relation between $logS_{23}$ and $log([OII]/[OIII])$ is readily apparent  in the lower panel of the figure. The peculiar position of IZw18 in the plots is 
due to its very low metallicity. 

Based on the above considerations, we have attempted a calibration of 
oxygen abundance through the sulphur abundance parameter $S_{23}$, using the 
data of Table 2 with the exclusion of the three objects for which no direct 
determinations of the electron temperature exist: S5 in M101, CC93 in M33 and 
the nucleus of NGC~3310. A linear fit to the data, taking into account the 
observational errors, give:
\[ 12+log(O/H) = 1.53 log S_{23} + 8.27 \]
with a correlation coefficient of 0.88 and a value of $\sigma$ = 0.15.

The data points corresponding to IZw18 deviate slightly from linearity. This 
is due to the high excitation of this object which increases the fraction of 
S$^{++}$ converted to S$^{3+}$ thus decreasing $S_{23}$.
A quadratical fit:
\[ 12+log(O/H) = 0.24 (log S_{23})^2 + 1.42 log S_{23} + 8.25 \]
provides a  better fit to these extremely low metallicity data.
The two fits are shown in the upper panel of Figure 3 by a solid and a dashed
line respectively. The corresponding residuals are shown in the lower panel as 
a function of the oxygen abundance. Solid dots correspond to the linear fit
while open circles correspond to the quadratic one. 

%%%%%%%%%%%%%%%%%%%%%%%%%%%%%%%%%%%%%%%%%%%%%%%%%%%%%%%%%%%%%%%%%%%%%%%%%%%%%%%%
%
%     Figure 4 Relation between O23 and S23.
%
%%%%%%%%%%%%%%%%%%%%%%%%%%%%%%%%%%%%%%%%%%%%%%%%%%%%%%%%%%%%%%%%%%%%%%%%%%%%%%%%

\begin{figure}
%\begin{minipage}{120mm}
\psfig{figure=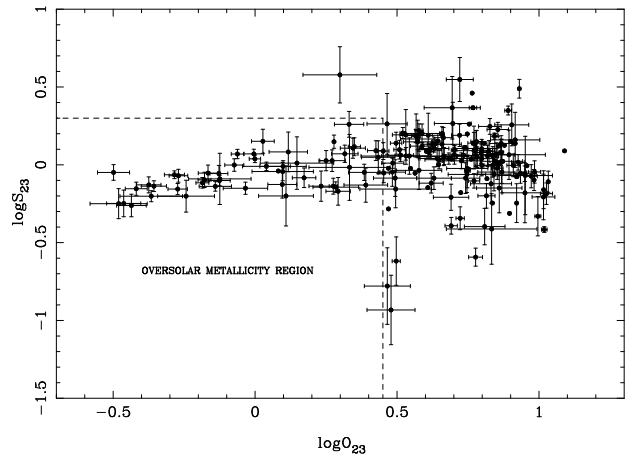,width=7cm,rheight=7.0cm,angle=270}
\caption{Relation between $O_{23}$ and $S_{23}$ for the objects in Table 1}
%\end{minipage}
\end{figure}

Figure 4 shows the relation between $S_{23}$ and $O_{23}$.
The largest value of $S_{23}$ corresponds to one of the observations of region
S5 in M101  ($S_{23}$ = 0.48), and the solar metallicity objects have $S_{23}$
between 0.28 and 0.48. On the other hand,  the object with the lowest
metallicity known, 
IZw18, has a value of $logO_{23}$= 0.47. These two facts taken together imply 
that objects with $logO_{23} \leq$ 0.47 and $-0.5 < logS_{23} \leq$ 0.28 will necessarily 
have oversolar abundances. These objects are  48 out of the 196 listed in 
Table 1 and therefore constitute about a quarter of the total sample. Most of them are circumnuclear star forming regions, HII regions in inner galactic discs and HII regions in Virgo cluster galaxies. 

For $logO_{23} > 0.45$ it can be seen that for the lowest metallicity objects 
there is a positive correlation between $O_{23}$ and $S_{23}$; NGC~5471, with 
$logO_{23} \simeq 1.00$ and $logS_{23} \simeq 0.10$, would lie at the end of
this correlation. Then, there is a trend of increasing $S_{23}$ with 
decreasing $O_{23}$ which corresponds to the "upper metallicity branch" of the 
$O_{23}$ abundance calibration. For values of $logO_{23}$ between 0.45 and 0.00,
the relation between the two abundance parameters is rather flat. Finally, for
values of $logO_{23} < 0.0$ a trend of decreasing $S_{23}$ with decreasing 
$O_{23}$ is apparent, indicating that, for the metallicities involved, the 
expected reversal of the $S_{23}$ {\it vs} metallicity relation has already 
taken place.  

%%%%%%%%%%%%%%%%%%%%%%%%%%%%%%%%%%%%%%%%%%%%%%%%%%%%%%%%%%%%%%%%%%%%%%%%%%%%%%%%
%
%     Figure 5 Abundance gradient in M~101
%
%%%%%%%%%%%%%%%%%%%%%%%%%%%%%%%%%%%%%%%%%%%%%%%%%%%%%%%%%%%%%%%%%%%%%%%%%%%%%%%%

\begin{figure}
%\begin{minipage}{120mm}
\psfig{figure=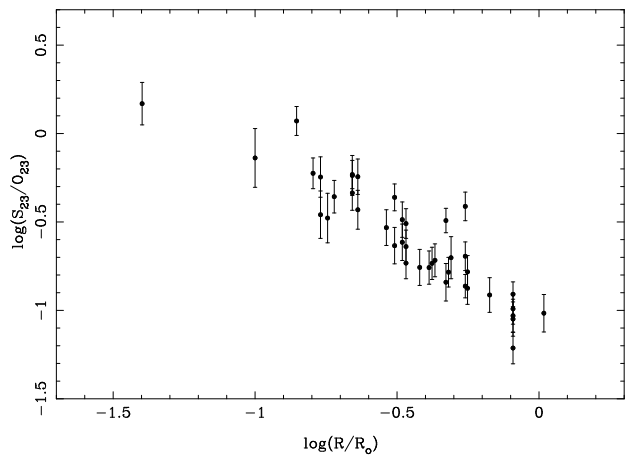,width=7cm,rheight=7.0cm,angle=270}
\caption{Variation of $log(S_{23}/O_{23})$ with galactic radius for M~101, which actally mimics a logarithmic abundance gradient through the disc.}
%\end{minipage}
\end{figure}

The behaviour of the $O_{23}$ and $S_{23}$ parameters is related with the 
different energies involved in the two sets of transitions and therefore a combination of both should, in principle, be a better indicator of metallicity 
than either of them alone. Figure 5 shows a plot of $log(S_{23}/O_{23})$ 
for the 
HII regions in M~101 (Kennicutt \& Garnett 1996). A single line logarithmic 
gradient is found. This might actually constitute a purely observational way to quantify galactic disc abundance gradients without the need to rely on 
theoretical photoionization models.

\section{Conclusions}

We have performed a new empirical calibration of nebular abundances using the 
sulphur abundance parameter $S_{23}$. This calibration is an alternative to 
the commonly used one based on the strong optical oxygen lines and presents 
several advantages. From the observational point of view, the lines are easily 
observable, both in low and high metallicity regions, and less affected by 
reddening. Furthermore, their intensities can be measured  
relative to nearby hydrogen recombination lines thus minimizing any effects 
due to uncertainties in flux calibration. On the theoretical side, their  
contribution to the cooling of the nebula  becomes important at  
electron temperatures lower than in the case of the traditional $O_{23}$ 
(previously called $R_{23}$) and therefore its relation with oxygen abundance 
remains single-valued up to metallicities close to solar. Also, the fact that 
$S_{23}$ is less dependent than $O_{23}$ on ionization parameter reduces the 
scatter in the relation.

The application of this new metallicity calibration can provide more accurate 
abundance determinations for objects with log$O_{23}$ between 0.5 and 1.2, 
{\it i. e.} oxygen abundances between 12+log(O/H)= 7.20 ($\simeq$ 0.02 times solar) and 12+log(O/H) = 8.80 ($\simeq$ 0.75 times solar). This is the range of 
metallicities found in HII galaxies and HII regions in irregular galaxies and  outer galactic discs.

Regarding HII regions of higher metallicity, the composed parameter 
$S_{23}/O_{23}$ can provide a better quantification of abundance gradients 
through galactic discs and might also hold the key to a future abundance 
calibration in this lower temperature regime.

\section*{Acknowledgements}

This work has been partially supported by DGICYT project PB-96-052. We would 
like to thank Elena Terlevich and Bernard Pagel for a careful reading of the 
manuscript.

\end{document}